\documentclass[preprint,onecolumn,12pt]{revtex4}
\usepackage{graphicx}
\usepackage{dcolumn}
\usepackage{bm}
\usepackage{amssymb}
\usepackage[colorlinks=true]{hyperref}
\usepackage{amsmath}
\usepackage{tikz}
\usetikzlibrary{calc,decorations.markings}
\DeclareMathOperator{\tr}{tr}
\usepackage{subfig}

\newcommand{\be}{\begin{equation}}
\newcommand{\ee}{\end{equation}}
\newcommand{\bea}{\begin{eqnarray}}
\newcommand{\eea}{\end{eqnarray}}
\newcommand{\bmx}{\begin{pmatrix}}
\newcommand{\emx}{\end{pmatrix}}

\newcommand{\gra}{\alpha}
\newcommand{\grb}{\beta}
\newcommand{\grg}{\gamma}
\newcommand{\grd}{\delta}
\newcommand{\gre}{\epsilon}

\newcommand{\grm}{\mu}

\newcommand{\grr}{\rho}
\newcommand{\grs}{\sigma}
\newcommand{\grt}{\tau}

\newcommand{\grf}{\phi}

\newcommand{\grc}{\psi}
\newcommand{\grw}{\omega}

\newcommand{\nn}{\nonumber}

\begin{document}

\title{Non-Markovian time evolution of an accelerated qubit}
\author{Dimitris Moustos}
\email{dmoustos@upatras.gr}
\author{Charis Anastopoulos}
\email{anastop@physics.upatras.gr}
\affiliation{Department of Physics, University of Patras, 26500 Patras, Greece}
\date{\today}

\begin{abstract}
We present a new method for evaluating  the response of  a moving qubit detector interacting with a scalar field in Minkowski spacetime.    We treat the detector as an open quantum system, but we do not invoke the Markov approximation. The evolution equations for the qubit density matrix are valid at all times, for all qubit trajectories, and they incorporate non-Markovian effects. We analyze in detail the case of uniform acceleration, providing a detailed characterization of all regimes where non-Markovian effects are significant. We argue that the most stable   characterization of acceleration temperature   refers to the {\em late time}   behavior of the detector because interaction with the field vacuum brings the qubit to a thermal state at the Unruh temperature. In contrast, the early-time transition rate, that is invoked in most discussions of acceleration temperature,  does not exhibit a thermal behavior when non-Markovian effects are taken into account. Finally, we note that the non-Markovian evolution derived here also applies to the mathematically equivalent problem of a static qubit interacting with a thermal field bath.
\end{abstract}

\maketitle

\pagebreak

\section{Introduction}
\subsection{Background}

A fundamental property of quantum field theory (QFT) in Minkowski spacetime is that all inertial observers agree on the number of particles in a given field state. This is not the case for non-inertial observers, because different non-inertial  observers define particles with respect to different field modes \cite{Full, Unruh2}.  The most important example is the Unruh effect \cite{Unruh}: for an observer moving with uniform proper acceleration $a$, the  Minkowski vacuum appears as a heat bath at the Unruh temperature $T_U = \frac{a}{2\pi}$.

In this paper, we develop a systematic method for describing the time evolution of microscopic accelerated detectors in Minkowski spacetime. Treating the detector as a two-level system (2LS), we derive the evolution equation for the associated reduced density matrix. Our method applies to all motions of the detector and incorporates effects beyond the often invoked Markov approximation.

QFT derivations of the Unruh effect employ the idealized notion of an eternally accelerated observer and they depend on global spacetime properties, such as the existence of a Rindler horizon \cite{Full2}. In order to demonstrate the physical relevance of the effect, it must be expressed in terms of   local physics \cite{Unruh, UW84, BL, AM, matsas, review}. The most commonly employed model to this end is the Unruh-DeWitt detector, introduced in Ref. \cite{Dewitt}.  An Unruh-DeWitt detector consists of a point-like quantum system, interacting through a monopole coupling with a quantum field and moving along a path in Minkowski spacetime.

The excitation rate of a moving detector is usually evaluated to leading order in perturbation theory \cite{Birrell}. For constant acceleration, the excitation rate is  constant and it depends on energy through the Planck distribution at the Unruh temperature.
 This feature is usually taken as a validation of the notion of acceleration temperature. However, the perturbative evaluation of the excitation rate has a restricted domain of applicability. It works best for {\em macroscopic detectors}, i.e., systems that leave a macroscopic record (a click) for every particle detection. For such detectors, the perturbative transition rate applies at all times, provided that the detector's temporal resolution is sufficiently large \cite{AnSav11, AnSav15}.  This is not the case for microscopic detectors, like for example, elementary particles, nuclei, or atoms. The leading-order perturbative evaluation of the transition rate applies only during very early times. It also ignores effects such as the back-action of the detector to the field and the spontaneous emission after excitations.  Taking such effects into account is important for understanding the physics of the acceleration temperature and for making concrete predictions about experiments that could measure    particle detection by moving detectors.

\subsection{Particle detectors as open quantum systems}

The Hamiltonian of a microscopic  two-level Unruh-DeWitt detector interacting with a quantum field is a special case of the spin-boson Hamiltonian \cite{spinboson}. This suggests the treatment of the Unruh-DeWitt detector as an open quantum system, with the quantum field playing the role of the environment \cite{BrePe07}. 

Open quantum systems are often described by  a Markovian master equation, in which the non-unitary terms are of second order to the system-environment coupling.
The second-order master equation is an approximation to the exact quantum dynamics. It is obtained at the limit where the system-environment coupling vanishes and the time parameter has been appropriately rescaled  \cite{Davies}. Alternatively, the master equation is derived from the successive implementation of  three  approximations \cite{BrePe07} :
\begin{itemize}
\item Born's approximation: For weak coupling between system and environment, the state of the environment is negligibly affected by the interaction with the system.
    \item The Markov approximation: The two-time correlation functions of the bath are approximated by delta functions.
  \item The (post-trace) Rotating Wave Approximation (RWA):  rapidly oscillating terms    in the interaction-picture evolution equation are ignored \cite{Agarwal, RWA}.
\end{itemize}
The second-order master equation is  an excellent approximation to a large class of problems. Nonetheless, it fails in many regimes. This can be seen by direct comparison with models in which exact solutions to the evolution equations are available, like, for example,  quantum Brownian motion models \cite{HPZ}. The main problem is the Markov approximation; it turns out to be too drastic for a large class of problems. Hence, the validity of the second-order master equation cannot be presupposed when studying the interaction between moving detectors and a quantum field, as in Ref. \cite{Benatti}. Indeed, if  the Unruh-DeWitt detector is modeled by a harmonic oscillator (instead of a qubit), the system is exactly solvable, and non-Markovian effects turn out to be significant \cite{Lin}.

\subsection{This work}
 In this article, we derive the evolution equations for a moving two-level detector interacting with a quantum scalar field  without invoking either the Markov approximation or the RWA. We do invoke the Born approximation, so our results are not exact.

 In principle, our method works for all possible trajectories of the 2LS in Minkowski spacetime. The results simplify significantly   for    trajectories characterized by static bath correlation functions \cite{LePf, Letaw}, like, for example, trajectories of uniform acceleration and of rotation with constant angular velocity. In this case, we obtain a closed expression for the   reduced density matrix as a function of time. Furthermore, we develop a systematic procedure for identifying the part of the time evolution that corresponds to Markovian dynamics and, consequently, for quantifying non-Markovian behavior.

Then, we analyze the case of uniform acceleration in full detail---an analogous analysis for other trajectory types will be undertaken in a different publication.   The effective dynamics   depends on the relative size of three parameters with the dimension of inverse time: the qubit's frequency $\omega$, the qubit's decay coefficient  $\Gamma_0$, and the acceleration $a$. We find that the Markov approximation works well in a large part of the parameters' space, but there are regimes in which non-Markovian effects are strong. In particular, this is the case for small accelerations $a < \omega$ or for very large accelerations $a >> \omega$.

 The most important conclusion from our analysis is that the relation between temperature and acceleration is best understood in the long-time limit. The qubit's asymptotic state turns out to be thermal at the Unruh temperature {\em even when the non-Markovian effects are taken into account}. In contrast, the Planckian form of the early time transition rate is valid only within the Markov approximation. For $a< \omega$, non-Markovian effects at early times imply a time-dependent transition rate with no clear relation to a Planckian spectrum. Thus, the detector's asymptotic state provides a more  
   fundamental and persistent characterization of the acceleration temperature. The field vacuum eventually brings any accelerated qubit it interacts with into a state of thermal equilibrium.

 The structure of the article is the following. In Sec. \ref{evoleq}, we derive the evolution equations for a moving qubit interacting with a quantum scalar field in Minkowski spacetime using only the Born approximation. In Sec. \ref{UAD}, we solve these equations for the case of a uniformly accelerated qubit. In Sec. \ref{impl}, we examine the physical implications of this solution. Finally, in Sec. \ref{concl}, we summarize and discuss our results.

 We work with units   $\hbar=c=k_{B}=1$.

\section{Time evolution of Unruh-DeWitt detectors }\label{evoleq}

An Unruh-DeWitt detector \cite{Dewitt,Birrell} is an ideal particle detector coupled to a quantum field with a monopole interaction and moving along a trajectory $x^{\mu}(\tau)$ in Minkowski spacetime, where $\tau $ is the proper time of the detector.

We model an Unruh-DeWitt detector by a 2LS of frequency $\grw$. The detector interacts with a massless scalar  field $\hat{\phi}$. The Hamiltonian of the combined system is
\be
\hat{H}=\hat{H}_0\otimes\hat{1}+\hat{1}\otimes\hat{H}_{\phi}+\hat{H}_{\text{int}},
\ee
where
\be
\hat{H_0}=\frac{\grw}{2}\hat{\grs}_3
 \ee
 is the 2LS Hamiltonian,
 \be
 \hat{H}_{\grf}=\int d^3x\left(\frac{1}{2}\hat{\pi}^2+\frac{1}{2}(\nabla \hat{\grf})^2\right)
  \ee
  is the Hamiltonian of the scalar field, 
  \be
  \hat{H}_{\text{int}}=g\hat{m}\otimes\hat{\phi}(\mathbf{x})
\ee
is the interaction Hamiltonian,  $g$ is the coupling constant, and $\hat{m} = \hat{\sigma}_1$ is the detector's monopole moment operator.

The evolution equation of the density matrix $\hat{\grr}_{\text{tot}}$ of the total system in the interaction picture is 
\be\label{von}
\frac{d}{d\grt}\hat{\grr}_{\text{tot}}(\grt)=-i\left[\hat{H}_\text{int}(\grt),\hat{\grr}_{\text{tot}}(\grt)\right],
\ee
where 
\be
\hat{H}_{\text{int}}(\grt)=g\hat{m}(\grt)\otimes\hat{\grf}\left[x^{\grm}(\grt)\right],
\ee
 with
\be
\hat{m}(\grt)=e^{i\grw\grt}\hat{\grs}_++e^{-i\grw\grt}\hat{\grs}_-
\ee
expressed in terms of the SU(2) ladder operators $\hat{\grs}_{\pm}$.

For weak system-field coupling, we solve Eq. (\ref{von}) using the Born approximation. We assume that the state
  of the total system at time $\grt$ approximates a tensor product
\be
\hat{\grr}_{\text{tot}}(\grt)\approx \hat{\grr}(\grt)\otimes\hat{\grr}_{\grf}(0),
\ee
where $\hat{\grr}$ is the reduced density matrix of the 2LS.

Then, Eq. (\ref{von}) becomes an integro-differential for the reduced density matrix $\hat{\grr}$  \cite{BrePe07}.
For a field  in its ground state $\hat{\grr}_{\grf}(0)=|0 \rangle\langle 0|$,  
\bea\label{dmevol}
\dot{\hat{\grr}}(\grt)=&&g^2\int_0^{\grt}ds\left[\hat{m}(s)\hat{\grr}(s)\hat{m}(\grt)-\hat{m}(\grt)\hat{m}(s)\hat{\grr}(s)\right]\Delta^+(\grt ;s)\nn \\&+&g^2\int_0^{\grt}ds\left[\hat{m}(\grt)\hat{\grr}(s)\hat{m}(s)-\hat{\grr}(s)\hat{m}(s)\hat{m}(\grt)\right]\Delta^-(\grt ;s),
\eea
where $\Delta^+(\grt ;s)=\langle 0|\hat{\grf}[x(\grt)]\hat{\grf}[x(s)]|0\rangle$ is the positive-frequency Wightman function and $\Delta^-(\grt ;s)=\langle 0|\hat{\grf}[x(s)]\hat{\grf}[x(\grt)]|0\rangle$ is the negative-frequency Wightman  function.

Expressing the density operator in a matrix form
\be\label{dmatrix}
\hat{\grr}(\grt)=
\bmx
\grr_{11}(\grt)&\grr_{10}(\grt)\\
\grr_{01}(\grt)&\grr_{00}(\grt)
\emx,
\ee
we obtain
\bea
\dot{\grr}_{11}(\grt) &=& -g^2\int_0^{\grt}ds\left[G^{++}(\grt; s)+G^{--}(\grt;s)\right]\grr_{11}(s)
\nonumber \\
&+& g^2\int_0^{\grt}ds\left[G^{+-}(\grt; s)+G^{-+}(\grt;s)\right]\grr_{00}(s), \label{dmb1}
\eea
\bea
\dot{\grr}_{00}(\grt) &=&-g^2\int_0^{\grt}ds\left[G^{+-}(\grt; s)+G^{-+}(\grt;s)\right]\grr_{00}(s) \nonumber \\
&+&g^2\int_0^{\grt}ds\left[G^{++}(\grt; s)+G^{--}(\grt; s)\right]\grr_{11}(s) ,  \label{dmb3}
\eea
\bea
\dot{\grr}_{10}(\grt)=&-&g^2\int_0^{\grt}ds\left[G^{++}(\grt; s)+G^{-+}(\grt; s)\right]\grr_{10}(s)\nn\\&+&g^2e^{2i\grw\grt}\int_0^{\grt}ds\left[G^{--}(\grt; s)+G^{+-}(\grt; s)\right]\grr_{01}(s) \label{dmb2},
\eea
where we defined the correlation functions
\bea\label{correl}
G^{++}(\grt; s) :=\Delta^+(\grt;s)e^{i\grw(\grt-s)}&,& \quad G^{--}(\grt;s) :=\Delta^-(\grt;s)e^{-i\grw(\grt-s)},\nn \\
 \quad G^{+-}(\grt; s) :=\Delta^+(\grt;s)e^{-i\grw(\grt-s)}&,& \quad G^{-+}(\grt; s) :=\Delta^-(\grt;s)e^{i\grw(\grt-s)}.
\eea

Equations (\ref{dmb1}--\ref{dmb2}) hold for any trajectory followed by the 2LS. They are derived using only the Born approximation.  We  used neither the  Markov approximation nor the RWA.

Equations (\ref{dmb1}--\ref{dmb2}) are most easily solved for static Wightman functions, i.e.,  $\Delta^{\pm} (\grt; s) = \Delta^{\pm} (\grt- s)$. This is possible for a specific class of spacetime trajectories \cite{LePf, Letaw},  which includes trajectories with constant proper acceleration and with rotation at constant angular velocity. For such trajectories, we express the correlation functions as \hspace{5cm} $G^{++}(\grt; s) =G^{++}(\grt - s), \ G^{--}(\grt; s) =G^{--}(\grt - s),\ G^{+-}(\grt; s) =G^{+-}(\grt - s)$ and $G^{-+}(\grt; s) =G^{-+}(\grt - s)$.  

We Laplace-transform  Eqs. (\ref{dmb1}--\ref{dmb2})  and use the convolution theorem to obtain
\bea\label{r11lp}
z\tilde{\grr}_{11}(z)-\grr_{11}(0)=-g^2\left(\tilde{G}^{++}(z)+\tilde{G}^{--}(z)\right)\tilde{\grr}_{11}(z)+g^2\left(\tilde{G}^{+-}(z)+\tilde{G}^{-+}(z)\right)\tilde{\grr}_{00}(z)
\eea
\bea\label{r00}
z\tilde{\grr}_{00}(z)-\grr_{00}(0)=-g^2\left(\tilde{G}^{+-}(z)+\tilde{G}^{-+}(z)\right)\tilde{\grr}_{00}(z)+g^2\left(\tilde{G}^{++}(z)+\tilde{G}^{--}(z)\right)\tilde{\grr}_{11}(z),
\eea
\bea\label{r10}
z\tilde{\grr}_{10}(z)-\grr_{10}(0)=&-&g^2\left(\tilde{G}^{++}(z)+\tilde{G}^{-+}(z)\right)\tilde{\grr}_{10}(z)\nn\\&+&g^2\left(\tilde{G}^{--}(z-2i\grw)+\tilde{G}^{+-}(z-2i\grw)\right)\tilde{\grr}_{01}(z-2i\grw),
\eea
where $\tilde{\grr}_{mn}(z)=\mathcal{L}\{\grr_{mn}(\grt)\}(z)$ is the Laplace transform of the density matrix  and $\tilde{G}^{ij}(z)$  is  the Laplace transform of the correlation functions;  $i,j = \pm$.

We solve Eqs. (\ref{r11lp}--\ref{r10})  for $\tilde{\grr}_{mn}(z)$,
\begin{eqnarray}
\tilde{\grr}_{11}(z) &=& \frac{[z+g^2\tilde{B}(z)]\grr_{11}(0) + g^2\tilde{B}(z) \grr_{00}(0) }{z\left[z+g^2\left(\tilde{A}(z)+\tilde{B}(z)\right)\right]},  \label{r11l}\\
 \tilde{\grr}_{00}(z) &=& \frac{ g^2\tilde{A}(z) \grr_{11}(0) +  [z+ g^2\tilde{A}(z)] \grr_{00}(0) }{z\left[z+g^2\left(\tilde{A}(z)+\tilde{B}(z)\right)\right]},  \label{r00l}\\
 \tilde{\grr}_{10}(z)&=&\frac{\left[z-2i\grw+ g^2\tilde{N}(z-2i\grw)\right]\grr_{10}(0)+g^2\tilde{N}(z-2i\grw)\grr_{01}(0)}{z(z-2i\grw)+g^2z\tilde{N}(z-2i\grw)) +g^2(z-2i\grw)\tilde{M}(z)}, \label{r10l}
\end{eqnarray}
where
\begin{eqnarray}
\tilde{A}(z) :=\tilde{G}^{++}(z)+\tilde{G}^{--}(z), \hspace{1cm} \tilde{B}(z) :=\tilde{G}^{+-}(z)+\tilde{G}^{-+}(z), \nonumber
  \\
\tilde{N}(z) :=\tilde{G}^{--}(z)+\tilde{G}^{+-}(z), \hspace{1cm} \tilde{M}(z) :=\tilde{G}^{++}(z)+\tilde{G}^{-+}(z). \label{correl2}
\end{eqnarray}
  From Eqs. (\ref{r11l}--\ref{r10l}), we determine $\rho_{mn}(\tau)$  by implementing the inverse Laplace transform, using the
  the Bromwich integral
\be
\grr(\grt)= \frac{1}{2\pi i}\int_{c-i\infty}^{c+i\infty}e^{zt}\tilde{\grr}(z)dz \label{bromw}
\ee
where $c$ is a real constant larger than the real part of all the singularities of the integrand \cite{Arf}.

\section{Uniformly accelerated qubit}\label{UAD}
In this section, we apply the results of Sec. \ref{evoleq} to the special case of
 a uniformly accelerated qubit detector. 
\subsection{The correlation functions}
We consider the hyperbolic trajectory
\be
x^0(\grt)=a^{-1}\sinh(a\grt), \quad x^1(\grt)=a^{-1}\cosh(a\grt), \quad x^2(\grt)=x^3(\grt)=0
\ee
where $a$ is the proper acceleration. The  corresponding Wightman functions are
\be\label{Wacc}
\Delta^{\pm}(\grt ;s)= -\lim_{\epsilon \rightarrow 0^+} \frac{a^2}{16\pi^2\sinh^2[a(\grt-s \mp i\gre)/2]}.
\ee

In what follows, we keep a finite value of $\epsilon$ in Eq. (\ref{Wacc}) and treat it as a regularization parameter. Thus, we will evaluate  the Laplace transforms $\tilde{G}^{ij}(z)$ of the correlations functions (\ref{correl}) for small but finite values of $\epsilon$.
Detailed  calculations are shown in Appendix \ref{LPT}. The result is 
\be
\tilde{A}(z)=\frac{1}{2}\tilde{\Sigma}(z)+\frac{\grw}{4\pi},
\ee
\be
\tilde{B}(z)=\frac{1}{2}\tilde{\Sigma}(z)-\frac{\grw}{4\pi},
\ee
where 
\bea\label{ltA}
\tilde{\Sigma}(z)=\tilde{A}(z)+\tilde{B}(z)=&-&\frac{a}{2\pi^2}-\frac{z-i\grw}{2\pi^2}\left[\log(e^{\grg}\gre a)+\grc\left(\frac{z-i\grw}{a}\right)\right]\nn\\&-&\frac{z+i\grw}{2\pi^2}\left[\log(e^{\grg}\gre a)+\grc\left(\frac{z+i\grw}{a}\right)\right]
\eea
and 
\be
\tilde{N}(z)=-\frac{a}{4\pi^2}-\frac{z+i\grw}{2\pi^2}\left[\log(e^{\grg}\gre a)+\grc\left(\frac{z+i\grw}{a}\right)\right],
\ee
\be\label{ltM}
\tilde{M}(z)=-\frac{a}{4\pi^2}-\frac{z-i\grw}{2\pi^2}\left[\log(e^{\grg}\gre a)+\grc\left(\frac{z-i\grw}{a}\right)\right];
\ee
$\grg$ is the Euler-Mascheroni constant and $\psi(x)$ is the digamma (psi) function \cite{Bateman,Abramowitz}.

\subsection{Evolution of the density matrix}

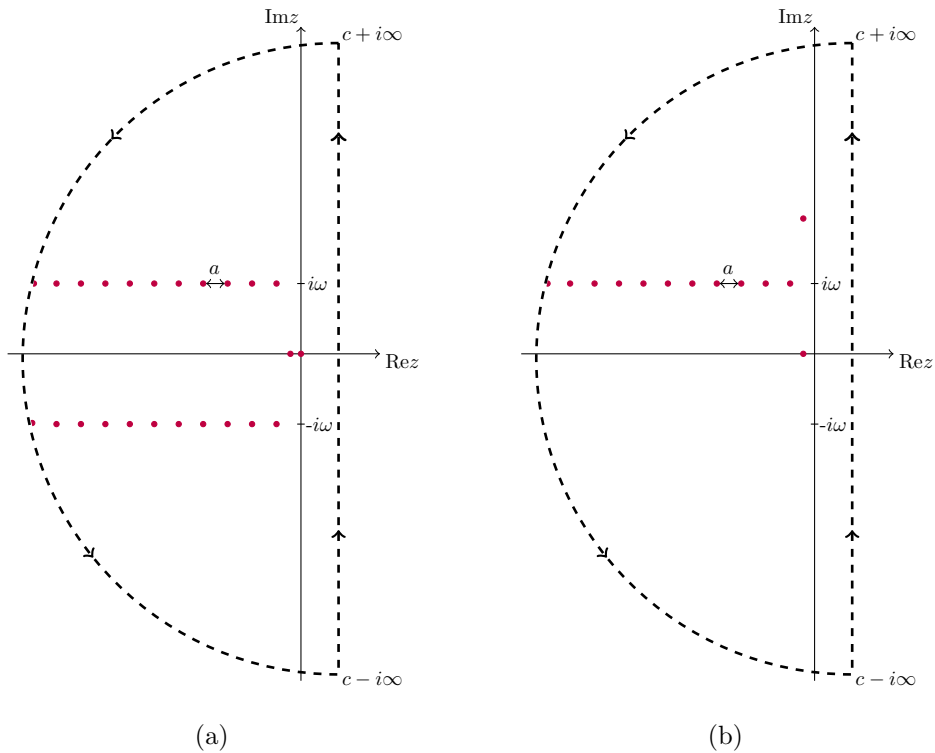
\begin{figure}
\subfloat[]{ \begin{tikzpicture}[scale=0.5, every node/.style={scale=0.7}]
\def\gap{0.2}
\def\bigradius{3}
\def\littleradius{0.5}

\draw [thin,->] (-2.6*\bigradius, 0) -- (0.7*\bigradius,0);
\draw [thin,->] (0, -2.9*\bigradius) -- (0, 2.9*\bigradius);

\node at (2.7,-0.2){$\text{Re}z$};
\node at (-0.5,9) {$\text{Im}z$};
\node at (0.45,1.87){$i\omega$};
\node at (0.5,-1.87){-$i\omega$};
\node at (1.9,8.5){$c+i\infty$};
\node at (1.9,-8.65){$c-i\infty$};

\draw[thin] (-0.1,1.87) -- (0.1,1.87);
\draw[thin] (-0.1,-1.87) -- (0.1,-1.87);

\draw [fill=purple,purple] (-0.65,1.87) circle [radius=0.07];
\draw [fill=purple,purple] (-1.3,1.87) circle [radius=0.07];
\draw [fill=purple,purple] (-1.95,1.87) circle [radius=0.07];
\draw [fill=purple,purple] (-2.6,1.87) circle [radius=0.07];
\draw [fill=purple,purple] (-3.25,1.87) circle [radius=0.07];
\draw [fill=purple,purple] (-3.9,1.87) circle [radius=0.07];
\draw [fill=purple,purple] (-4.55,1.87) circle [radius=0.07];
\draw [fill=purple,purple] (-5.2,1.87) circle [radius=0.07];
\draw [fill=purple,purple] (-5.85,1.87) circle [radius=0.07];
\draw [fill=purple,purple] (-6.5,1.87) circle [radius=0.07];

\draw [fill=purple,purple] (-0.65,-1.87) circle [radius=0.07];
\draw [fill=purple,purple] (-1.3,-1.87) circle [radius=0.07];
\draw [fill=purple,purple] (-1.95,-1.87) circle [radius=0.07];
\draw [fill=purple,purple] (-2.6,-1.87) circle [radius=0.07];
\draw [fill=purple,purple] (-3.25,-1.87) circle [radius=0.07];
\draw [fill=purple,purple] (-3.9,-1.87) circle [radius=0.07];
\draw [fill=purple,purple] (-4.55,-1.87) circle [radius=0.07];
\draw [fill=purple,purple] (-5.2,-1.87) circle [radius=0.07];
\draw [fill=purple,purple] (-5.85,-1.87) circle [radius=0.07];
\draw [fill=purple,purple] (-6.5,-1.87) circle [radius=0.07];
\draw [fill=purple,purple] (0,0) circle [radius=0.07];
\draw [fill=purple,purple] (-0.28,0) circle [radius=0.07];

\draw [fill=purple,purple] (-7.10,1.94) arc (90:-110:0.07);
\draw [fill=purple,purple] (-7.18,-1.78) arc (120:-90:0.07);

\draw[thin,<->](-2.03,1.87) -- (-2.52,1.87);
\node at (-2.3,2.25){$a$};

 \draw[dashed,line width=1pt,yshift=93pt,decoration={ markings,
  mark=at position 0.156 with {\arrow[line width=1pt]{>}},mark=at position 0.438 with {\arrow[line width=1pt]{>}},mark=at position 0.70 with {\arrow[line width=1pt]{>}},mark=at position 0.945 with {\arrow[line width=1.2pt]{>}}},postaction={decorate}]
 (1,5) arc (90:270:8.4cm)-- (1,5);
\end{tikzpicture}}
\hspace{1cm}
  \subfloat[]{ \begin{tikzpicture}[scale=0.5, every node/.style={scale=0.7}]
\def\gap{0.2}
\def\bigradius{3}
\def\littleradius{0.5}

\draw [thin,->] (-2.6*\bigradius, 0) -- (0.7*\bigradius,0);
\draw [thin,->] (0, -2.9*\bigradius) -- (0, 2.9*\bigradius);

\node at (2.7,-0.2){$\text{Re}z$};
\node at (-0.5,9) {$\text{Im}z$};
\node at (0.45,1.87){$i\omega$};
\node at (0.5,-1.87){-$i\omega$};
\node at (1.9,8.5){$c+i\infty$};
\node at (1.9,-8.65){$c-i\infty$};

\draw[thin] (-0.1,1.87) -- (0.1,1.87);
\draw[thin] (-0.1,-1.87) -- (0.1,-1.87);

\draw [fill=purple,purple] (-0.65,1.87) circle [radius=0.07];
\draw [fill=purple,purple] (-1.3,1.87) circle [radius=0.07];
\draw [fill=purple,purple] (-1.95,1.87) circle [radius=0.07];
\draw [fill=purple,purple] (-2.6,1.87) circle [radius=0.07];
\draw [fill=purple,purple] (-3.25,1.87) circle [radius=0.07];
\draw [fill=purple,purple] (-3.9,1.87) circle [radius=0.07];
\draw [fill=purple,purple] (-4.55,1.87) circle [radius=0.07];
\draw [fill=purple,purple] (-5.2,1.87) circle [radius=0.07];
\draw [fill=purple,purple] (-5.85,1.87) circle [radius=0.07];
\draw [fill=purple,purple] (-6.5,1.87) circle [radius=0.07];

\draw [fill=purple,purple] (-0.3,3.6) circle [radius=0.07];
\draw [fill=purple,purple] (-0.3,0) circle [radius=0.07];

\draw [fill=purple,purple] (-7.10,1.94) arc (90:-110:0.07);

\draw[thin,<->](-2.03,1.87) -- (-2.52,1.87);
\node at (-2.3,2.25){$a$};

 \draw[dashed,line width=1pt,yshift=93pt,decoration={ markings,
  mark=at position 0.156 with {\arrow[line width=1pt]{>}},mark=at position 0.438 with {\arrow[line width=1pt]{>}},mark=at position 0.70 with {\arrow[line width=1pt]{>}},mark=at position 0.945 with {\arrow[line width=1.2pt]{>}}},postaction={decorate}]
 (1,5) arc (90:270:8.4cm)-- (1,5);
\end{tikzpicture}}
  \caption{Bromwich contour and poles of (a) the Laplace-transformed diagonal elements  and (b) the $\tilde{\rho}_{10}(z)$ element  of the density matrix. Integration is along a straight line from $c-i\infty$ to $c+i\infty$, where $c$ is  a real constant larger than the real part of the
poles of the integrand. The contour is closed by a semicircle of radius $R\to \infty$. }
\label{poles}
\end{figure}

The digamma function $\psi(z)$ that appears in   Eqs. (\ref{ltA}---\ref{ltM}) is a meromorphic function with simple poles at $z = -n, n=0, 1, 2 \ldots$. This implies that the Bromwich integral (\ref{bromw}) can be evaluated on a contour that encloses all poles of the integrand as in Fig. \ref{poles}. In Fig. \ref{poles}, we also show the poles of the 
  Laplace-transformed elements $\tilde{\rho}_{mn}(z)$ of the density matrix.

\subsubsection{Diagonal elements}

  The diagonal elements have two poles, one at $z = 0$ and one  at $z = - \Gamma +O(g^4)$, where
 \begin{eqnarray}
 \Gamma =  \Gamma_0   \coth\left(\frac{\pi\grw}{a}\right) \label{decaya}
\end{eqnarray}
is expressed in terms of the decay constant of a static qubit
\be
\Gamma_0 = \frac{g^2\grw}{2\pi}.
\ee
The relevant calculations are presented in Appendix \ref{Markopolo}.

The diagonal terms also have two infinite sequences of poles at
\be
z = -na\pm i\grw-\frac{g^2a}{2\pi^2}\frac{1}{1\mp\frac{i\grw}{na}} + O(g^4), \label{seqdiag}
\ee
where $n = 1, 2, 3, \ldots$. 

The second-order master equation for  this system considers only the poles at $z = 0$ and $z = -\Gamma$, which lie within a distance of order $O(g^2)$ from the double pole at $z = 0$ of the unperturbed propagator. The contribution of the poles (\ref{seqdiag}) is ignored. The latter poles characterize the deviation of the system from the Markovian behavior, and for this reason, we refer to them (and to their counterparts for the off-diagonal elements) as {\em non-Markovian poles}.

By means of Jordan's lemma, the integral along the semicircle in Fig.  \ref{poles}  vanishes, and the Bromwich integral is evaluated employing the residue theorem. We obtain
\begin{eqnarray}
\grr_{11}(\grt)&=&\frac{1}{2}\left(1-\frac{\Gamma_0}{\Gamma}\right)+\frac{\Gamma_0}{2\Gamma}e^{-\Gamma \grt}-\frac{1}{2}\left[e^{-\Gamma \grt}+\frac{2\Gamma_0}{ \grw}S_1(\grw;a;\grt)\right](\grr_{00}(0)-\grr_{11}(0)) \label{r11}\\
\grr_{00}(\grt) &=& 1 - \grr_{11}(\grt),
\end{eqnarray}
where the function $S_1(\grw;a;\grt)$ incorporates the contribution of the non-Markovian poles. Here, $S_1$ vanishes at $\tau = 0$, it exhibits a jolt at very early times and then becomes, to  leading-order in $g^2$,
\bea
S_1(\grw;a;\grt)&=&\frac{1}{\pi}Re\left\{\sum_{n=1}^{\infty}\frac{ne^{-na\grt}}{\left(n-i\frac{\grw}{a}\right)^2}\right\}\nn\\
&=& \frac{e^{-a\grt}}{\pi} Re\left\{e^{i\grw\grt}\left[\Phi\left(e^{-a\grt},1,1-\frac{i\grw}{a}\right)+\frac{i\grw}{a}\Phi\left(e^{-a\grt},2,1-\frac{i\grw}{a}\right)\right]\right\}, \label{S1}
\eea
where $\Phi(w,s,\gra)$ is the Lerch transcendent \cite{Ryzhik}.

Note that Eq. (\ref{S1}) does not apply to very early times ($\tau < O(g^4)$),   because it diverges logarithmically as $\tau \rightarrow 0$,
\be
 S_1(\grw;a;\grt) \simeq - \log(a \tau).
\ee
 Such logarithmic divergences appear in analogous calculations of non-Markovian effects   \cite{Lin, Svaiter}. In the present context, they arise because the perturbative evaluation of the non-Markovian poles, Eq. (\ref{seqdiag}), fails at very early times.

\subsubsection{Off-Diagonal elements} 

The Laplace transform of the off-diagonal element $\tilde{\grr}_{10}(z)$ has two poles, one   at
\be
z = -i(C +\Delta\grw)-\frac{\Gamma}{2} + O(g^4),
\ee
and one at
\be
z = 2i\grw+i (C +\Delta\grw)-\frac{\Gamma}{2} + O(g^4).
\ee
In the above expressions,
\be
C = \frac{\Gamma_0}{\pi}\log(e^{\grg}\gre\grw)
\ee
is an acceleration independent frequency renormalization term that diverges logarithmically as $\epsilon \rightarrow 0^+$, and
\be\label{Lamb}
\Delta\grw=\frac{\Gamma_0}{\pi}\left[\log(a/\grw )+Re\left\{\grc\left(\frac{i\grw}{a}\right)\right\}\right]
\ee
is the {\em finite} Lamb-shift of the frequency due to acceleration.

 The density matrix element $\tilde{\rho}_{10}(z)$ also has an infinite sequence of non-Markovian poles at
\be
z=-na+i\grw-\frac{g^2 a}{\pi^2}\frac{1}{1+(\frac{\grw}{na})^2},
\ee
where $n = 1, 2, 3, \ldots$.

We evaluate  the Bromwich integral and switch back to the Schr\"odinger picture to obtain
\bea
\grr_{10}(\grt)&=& \frac{\omega}{\bar{\omega}} e^{-i \bar{\omega}\grt -\frac{\Gamma}{2}\grt}\grr_{10}(0) - \frac{\Gamma_0}{\omega} S_2(\grw;a;\grt)\grr_{10}(0) + \frac{2\Gamma_0}{\omega} S_3(\grw;a;\grt)\grr_{10}(0) - \frac{\Gamma_0}{\omega} S_2(\grw;a;\grt)\grr_{01}(0)
 \nn\\
&& - \frac{\bar{\omega}-\omega}{\bar{\omega}}  e^{-\frac{\Gamma}{2} \grt}\cos(\bar{\omega} \tau) \grr_{01}(0) +\frac{\Gamma}{2 \bar{\omega}}e^{-\frac{\Gamma}{2} \grt} \sin(\bar{\omega} \tau)\grr_{01}(0),
\eea
where we defined the shifted frequency as 
\be
\bar{\omega} = \omega + C + \Delta \omega.
\ee
 The functions $S_2$ and $S_3$ contain the contribution of the non-Markovian poles. To leading-order in $g^2$, they read
\be\label{S2}
S_2(\grw;a;\grt)=\frac{1}{\pi} \sum_{n=1}^{\infty}e^{-na\grt}\frac{n}{n^2+(\frac{\grw}{a})^2}
\ee
and
\be
S_3(\grw;a;\grt)= \frac{1}{\pi} \sum_{n=1}^{\infty}e^{-na\grt}\frac{n^2(n+i \grw/a)}{\left(n^2+(\frac{\grw}{a})^2\right)^2}.
\ee

\subsection{Degenerate qubit: $\grw=0$}

 If $\grw = 0$,  the diagonal elements  (\ref{r11l})--(\ref{r00l}) of the density matrix  have two poles, one at $z = 0$ and one at $z = - \Gamma' +O(g^4)$, where
 \be
 \Gamma'=\frac{g^2a}{2\pi^2},
 \ee
 is the decay constant (\ref{decaya}) at the limit $ a >> \omega$. 
 
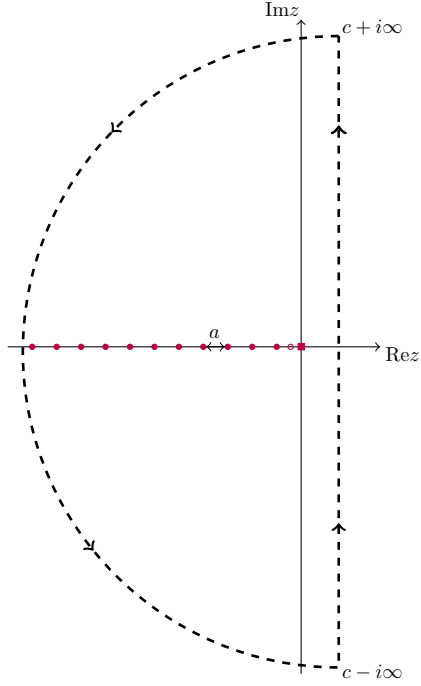
\begin{figure}
\begin{tikzpicture}[scale=0.5, every node/.style={scale=0.7}]
\def\gap{0.2}
\def\bigradius{3}
\def\littleradius{0.5}

\draw [thin,->] (-2.6*\bigradius, 0) -- (0.7*\bigradius,0);
\draw [thin,->] (0, -2.9*\bigradius) -- (0, 2.9*\bigradius);

\node at (2.7,-0.2){$\text{Re}z$};
\node at (-0.5,9) {$\text{Im}z$};

\node at (1.9,8.5){$c+i\infty$};
\node at (1.9,-8.65){$c-i\infty$};

\draw [fill=purple,purple] (-0.65,0) circle [radius=0.07];
\draw [fill=purple,purple] (-1.3,0) circle [radius=0.07];
\draw [fill=purple,purple] (-1.95,0) circle [radius=0.07];
\draw [fill=purple,purple] (-2.6,0) circle [radius=0.07];
\draw [fill=purple,purple] (-3.25,0) circle [radius=0.07];
\draw [fill=purple,purple] (-3.9,0) circle [radius=0.07];
\draw [fill=purple,purple] (-4.55,0) circle [radius=0.07];
\draw [fill=purple,purple] (-5.2,0) circle [radius=0.07];
\draw [fill=purple,purple] (-5.85,0) circle [radius=0.07];
\draw [fill=purple,purple] (-6.5,0) circle [radius=0.07];
\draw [fill=purple,purple] (-7.15,0) circle [radius=0.07];

\draw [fill=purple,purple] (-0.08,-0.08) rectangle (0.08,0.08);
\draw [purple] (-0.28,0) circle [radius=0.07];

\draw[thin,<->](-2.03,0) -- (-2.52,0);
\node at (-2.3,0.35){$a$};

 \draw[dashed,line width=1pt,yshift=93pt,decoration={ markings,
  mark=at position 0.156 with {\arrow[line width=1pt]{>}},mark=at position 0.438 with {\arrow[line width=1pt]{>}},mark=at position 0.70 with {\arrow[line width=1pt]{>}},mark=at position 0.945 with {\arrow[line width=1.2pt]{>}}},postaction={decorate}]
 (1,5) arc (90:270:8.4cm)-- (1,5);
\end{tikzpicture}
 \hspace{0.8cm}
\caption{Bromwich contour and poles of both diagonal and off-diagonal elements of the density matrix in the degenerate qubit limit.}
  \label{degencont}
\end{figure}

 The diagonal elements also have 
  an infinity sequence of poles at
\be
z=-na-\frac{g^2a}{\pi^2}+O(g^4),
\ee
as shown in Fig. \ref{degencont}. We evaluate the Bromwich integral (\ref{bromw}) to obtain 
\be
\grr_{11}(\grt)=\frac{1}{2}-\frac{1}{2}\left[e^{-\Gamma'\grt}-\frac{2\Gamma'}{a}\log(1-e^{-a\grt})\right]
(\grr_{00}(0)-\grr_{11}(0)). \label{degen}
\ee

The off-diagonal elements,  Eq. (\ref{r10l}), have the same poles with the diagonal elements. Calculating the Bromwich integral, we find
\bea
\grr_{10}(\grt)
&=&\frac{1}{2}\left(\grr_{10}(0)+\grr_{01}(0)\right)+
\frac{1}{2}e^{-\Gamma'\grt}\left(\grr_{10}(0)-\grr_{01}(0)\right)\nn\\
&&
+\frac{\Gamma'}{a}\log\left(1-e^{-a\grt}\right)\left(\grr_{10}(0)+\grr_{01}(0)\right)-\frac{2\Gamma'}{a}
\log\left(1-e^{-a\grt}\right)\grr_{10}(0). \label{deg01}
\eea

\section{Implications}\label{impl}

\subsection{Markovian regime and thermal behavior}
The second-order master equation for open quantum systems is derived using the Markov approximation, which involves the substitution of the bath  correlation functions with delta functions. This procedure is roughly equivalent with the approximation of keeping only the contribution from the Markovian poles in the evolution equations derived in Sec. \ref{UAD}. Then,
the diagonal elements coincide with the ones obtained from the second-order master equation
\be
\grr_{11}(\grt)=\frac{1}{2}\left(1-\frac{\Gamma_0}{\Gamma}\right)+\frac{\Gamma_0}{2\Gamma}e^{-\Gamma \grt}-\frac{1}{2} e^{-\Gamma \grt}  (\grr_{00}(0)-\grr_{11}(0)) \label{markov}.
\ee
Equation (\ref{markov})   exhibits the thermal behavior characteristic of the Unruh effect, both at early and late times.

At early times ($\Gamma t << 1$), and for $\rho_{11}(0) = 0$, Eq. (\ref{markov}) implies that
\be
\grr_{11}(\grt) =\Gamma_0\frac{\grt}{e^{\frac{2\pi\grw}{a}}-1}; \label{thrmtr}
\ee
i.e., the transition rate to the excited state is constant,
\begin{eqnarray}
w = \frac{\Gamma_0}{e^{\frac{2\pi\grw}{a}}-1} \label{trr}
 \label{tranw}
 \end{eqnarray}
and identical to the transition  rate of  a static qubit in a thermal bath at the Unruh temperature  $T_U = \frac{a}{2\pi}$.

The rate derived here coincides with the rate commonly obtained through {\em first order} perturbation theory \cite{Dewitt, Birrell}. The same expression  applies to macroscopic detectors \cite{AnSav11} at all times. However, for microscopic probes, such as the qubit considered in this paper, Eq. (\ref{trr})  applies only at  times $t << \Gamma^{-1}$.

In the long-time limit ($\Gamma t >> 1$), the density matrix approaches the equilibrium value
\be
\hat{\rho}_{\infty} = \left( \begin{array}{cc} \frac{e^{-\frac{2\pi\grw}{a}}}{e^{-\frac{2\pi\grw}{a}}+1}& 0 \\0 &\frac{1}{e^{-\frac{2\pi\grw}{a}}+1} \end{array} \right), \label{asympt}
\ee
which is a thermal density matrix at temperature $T_U$. Equation (\ref{asympt}) applies even when we keep the contribution from the non-Markovian poles. In our opinion, this is a much stronger manifestation of thermal behavior than the transition rate (\ref{tranw}). The accelerated qubit experiences the field vacuum as a genuine thermal bath and eventually settles at a thermal state.

\subsection{Lamb shift}
Another indication of thermal behavior comes from the
 Lamb shift, Eq. (\ref{Lamb}), written  in terms of the Unruh temperature as
\be\label{LambU}
\Delta\grw=\frac{\Gamma_0}{\pi}\left[\log\left(2\pi \frac{T_U}{\grw} \right)+Re\left\{\grc\left(\frac{i\grw}{2\pi T_U}\right)\right\}\right].
\ee

We compare this expression with the frequency shift induced by a thermal field bath for a static qubit.  In the Born-Markov approximation  (see, Ref. \cite{Carmichael} and also Appendix \ref{appme})
\be\label{mastershift}
\Delta\grw=\frac{\Gamma_0}{\grw\pi}PV\int_0^{\infty}d\grw_{\mathbf{k}}\frac{\grw_{\mathbf{k}}}{e^{\frac{\grw_{\mathbf{k}}}{T_U}}-1}\left(\frac{1}{\grw_{\mathbf{k}}+\grw}-\frac{1}{\grw_{\mathbf{k}}-\grw}\right),
\ee
where $\grw_{\mathbf{k}}$ is the frequency of the scalar field mode with momentum $\mathbf{k}$  and $PV$ indicates the Cauchy principal value.  We evaluated Eq. (\ref{mastershift}) numerically  and found  that it coincides with Eq.  (\ref{LambU}). Thus, the Lamb shift due to acceleration coincides with that of a thermal bath. (Note that our result differs by a multiplicative factor from that of Ref.    \cite{Audretsch}).

\subsection{Non-Markovian effects}\label{NonMark}
An important advantage of the method developed here is that it works beyond the usual Markov approximation and that it provides a precise quantification of  non-Markovian effects. These effects arise from the contribution of the non-Markovian poles to the solutions of the time evolution equations.

\begin{figure}
\subfloat[$\grw/a=1.1$]{\includegraphics[scale=0.4]{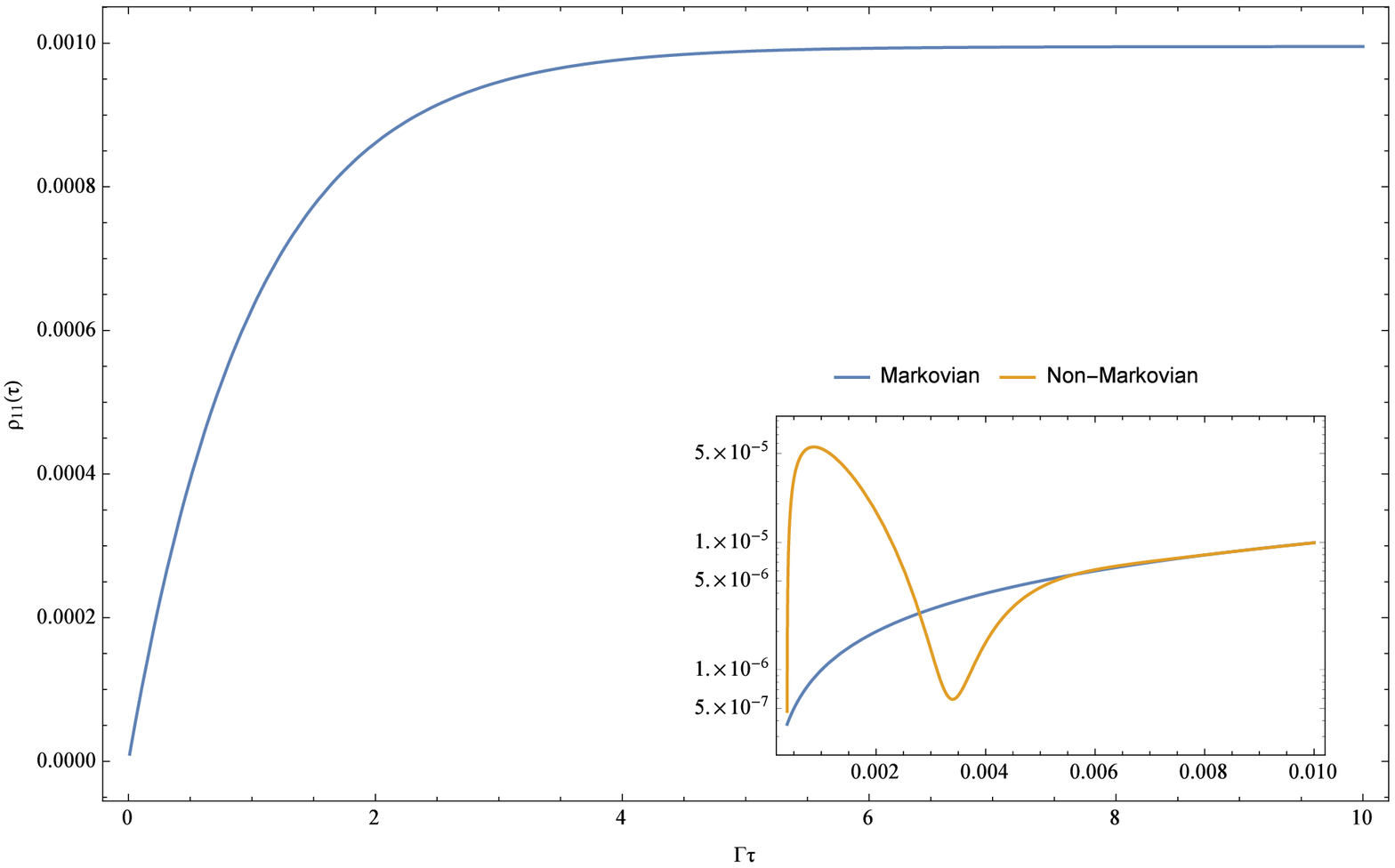}}  \hspace{1cm}
\subfloat[$\grw/a=0.9$]{\includegraphics[scale=0.4]{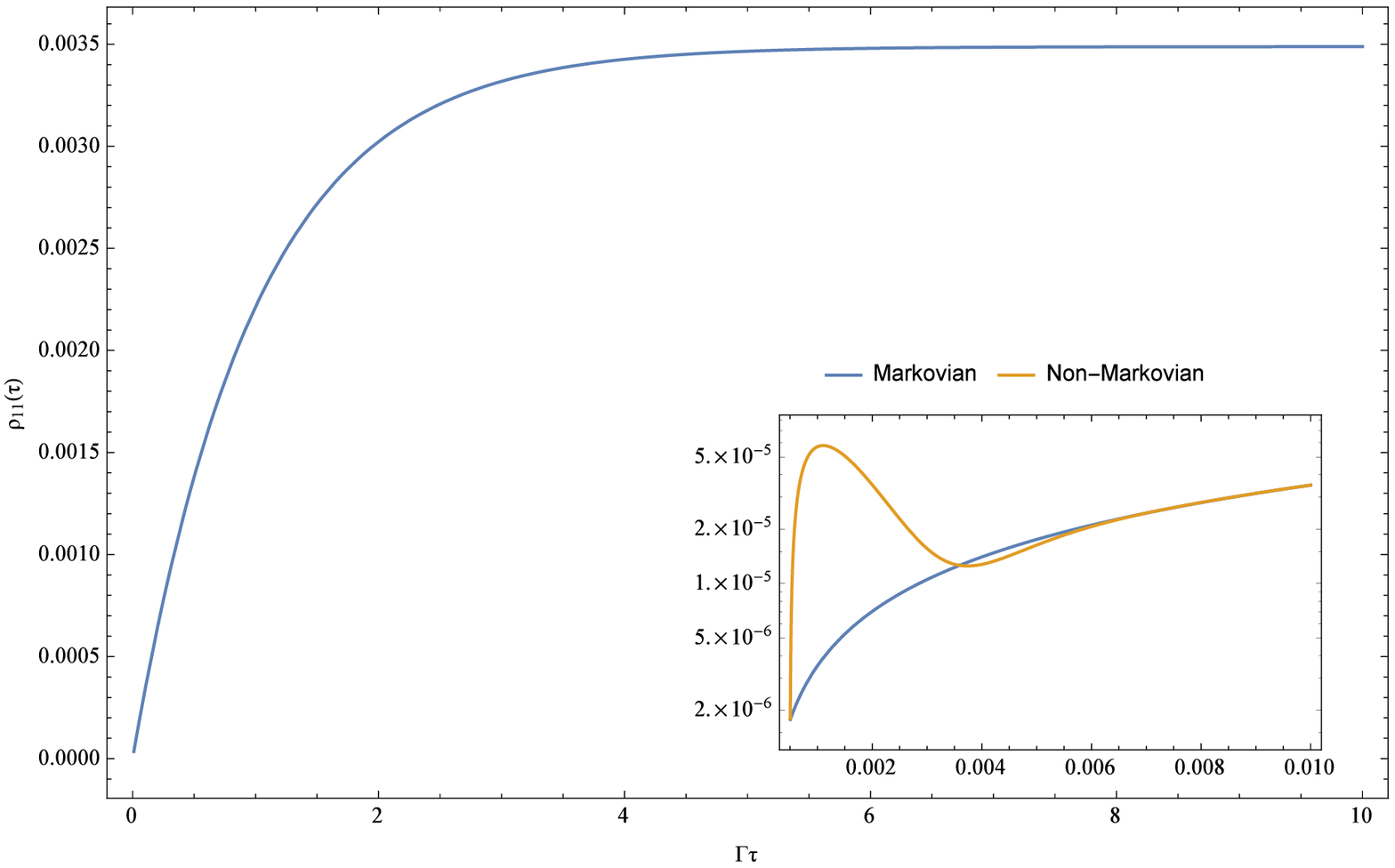}} \hspace{1cm}
\subfloat[$\grw/a=0.5$]{\includegraphics[scale=0.4]{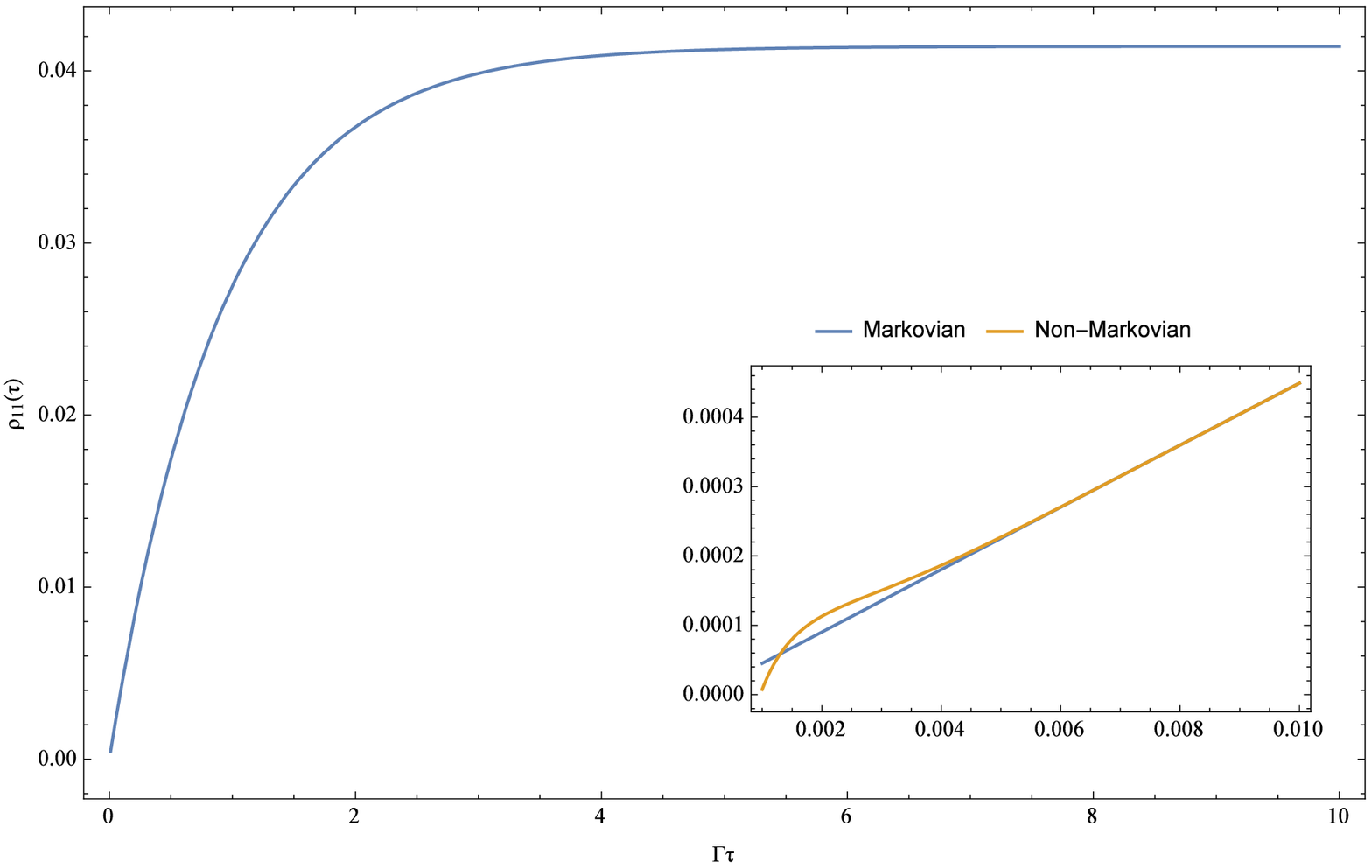}}
\caption{Time evolution of the $\grr_{11}(\grt)$ element of the density matrix for different values of $\grw/a$, when $\Gamma_0/\grw=10^{-3}$ and the qubit is initially found in its ground state $\grr_{00}(0)=1$. Evolution at early times is shown in the inserted plots.}
\label{evol}
\end{figure}

Non-Markovian effects are significant at early times ($\grt$ of order $\omega^{-1}$),  and they are particularly pronounced when the acceleration $a$ is smaller than the qubit's frequency $\omega$. This is shown in Fig. \ref{evol}, where the occupation probability of the excited state $\rho_{11}(\grt)$ is plotted as a function of a dimensionless time-parameter for an initial state with $\rho_{11}(0) = 0$. For $a$ close to $\omega$ or smaller, the oscillations induced by the non-Markovian terms dominate over Markovian ones. This effect diminishes with increasing $a$ and is practically absent for $a \gtrapprox 2 \omega$.

An immediate consequence  is that the identification of thermal behavior by the transition rate at early times is an artifact of the approximation employed, either  perturbation theory to the lowest order, or the Markov approximation. Measurements of the transition rate would not find a excitation rate of the form (\ref{thrmtr}) with an obvious thermal interpretation, unless $a$ is significantly larger than $\omega$. This implies that the defining feature of thermal behavior for accelerated qubits is the convergence to a thermal state at the long-time limit, and not the early-time transition rate, as is generally assumed. This conclusion applies only to microscopic probes---for macroscopic detectors the expression from perturbation theory is sufficient at all times \cite{AnSav11}.

At later times, the Markov approximation is valid, except for the limit  of very weak    acceleration. To see this, consider Eq. (\ref{r11}). At long times, the deviation of the Markovian terms from the equilibrium solution is of the order of $e^{-\Gamma \grt}$, while the non-Markovian terms decay within a time scale of $\alpha^{-1}$. Thus, if $\alpha < \Gamma$, convergence to equilibrium is dominated by the non-Markovian terms, and the relaxation time  is of the order of  $a^{-1}$ rather than $\Gamma^{-1}$.

The degenerate qubit limit, Eq. (\ref{degen}), applies to the regime of ultra-strong accelerations. In this regime, the
Markov approximation   breaks down at all   times. The asymptotic state is independent of the Markov approximation, as it is not affected by non-Markovian terms.

There are also significant qualitative differences between our results for the evolution of an accelerated qubit and the exact results of Ref. \cite{Lin} for an accelerated harmonic oscillator. In Ref. \cite{Lin}, the  Planckian transition probability (\ref{trr}) is recovered only in the limits of either ultra-high acceleration $a>>\grw$ or ultra-weak coupling $\grw>>\Gamma_0$,  as long as $a^{-1}<<\grt<<\Gamma_0^{-1}$.  Both limits correspond to the Markovian regime. Furthermore, in Ref. \cite{Lin} , non-Markovian effects can result in late-time detector readings that are totally different from the Unruh temperature, while we found that the long-time limit of the qubit detector is always a thermal state.

\subsection{Decoherence}
The field environment decoheres any initial superposition state of the form $c_1|0\rangle +c_2|1\rangle$, as seen in the evolution    of the off-diagonal element of the density matrix, $\rho_{10}(\grt)$.
 If we keep only the contribution of  the Markovian pole, we find that the decoherence rate equals $\frac{\Gamma}{2}$. This rate coincides with the decoherence rate  of a static qubit in a thermal bath at the Unruh temperature.

 We note that we cannot construct the Markovian time evolution by keeping only the contribution of the Markovian poles in the evolution of $\rho_{10}(\grt)$: the resulting expression does not have the correct behavior at $\grt = 0$. This is unlike the case of the diagonal elements considered earlier. The derivation of a  Markovian master equation that also describes the off-diagonal terms involves the more drastic step of modifying the bath correlation functions.

We found that the contribution of the non-Markovian terms to  $\rho_{10}(\grt)$ is small. In general, it does not lead to significant qualitative changes of the the decoherence process.  There are two exceptions. The first is the
   ultra-weak acceleration regime, where the asymptotic decoherence rate is $a$ rather than $\frac{\Gamma}{2}$. However,  this occurs at times so late that most of the initial phase information has already been lost.

   The second exception is the regime of ultra-strong acceleration, where  Eq. (\ref{deg01}) for the degenerate qubit applies. In this regime, the Markov approximation breaks at all times. We also note that there is no decoherence in the degenerate qubit limit; the asymptotic state preserves a substantial part of the initial phase information. The preservation of coherence applies only for an exactly degenerate qubit. For  approximately degeneracy, the off-diagonal elements are eventually suppressed, but the decoherence rate is very small.

\subsection{Mathematically equivalent field baths}

The evolution equations for the reduced density matrix, derived in Sec. \ref{UAD}, apply to all qubits interacting with a field bath characterized by correlation functions equivalent to (\ref{Wacc}).  In particular, they apply   to (i) static qubits in a thermal field bath and (ii) comoving qubits in de-Sitter spacetime.

\subsubsection{Thermal field bath} 
First, we consider the positive-frequency thermal Wightman function for a massless scalar field
\be
\Delta^+_{\grb}(\grt,\mathbf{x};\grt +s,\mathbf{x}) =-\frac{1}{4\grb^2\sinh^2\left[\pi\left(s-i\gre\right)/\grb\right]}-\frac{i}{4\pi}\sum_{n=-\infty}^{\infty}\grd\left(s+in\grb\right), \label{thermalw}
\ee
 where the $n=0$ is excluded from the summation.
 The first term in the r.h.s. of Eq. (\ref{thermalw}) coincides with the Wightman functions (\ref{Wacc}). Viewed as a function of $s$, the second term has support along a discrete set of points on the imaginary axis. It does not contribute to the Laplace transform of  $\Delta^+_{\grb}$ because the latter involves integration over the real axis of $s$. Thus, the correlation functions (\ref{Wacc}) for the accelerated qubit in vacuum and the static qubit in a thermal bath coincide for $\beta = T_U^{-1}$.

Consequently, our method describes non-Markovian processes for qubits coupled to a thermal field bath---see, for example,  \cite{Shresta04, Shibata, MaPe}. In particular, the results of Sec. \ref{NonMark} imply the following.
 \begin{enumerate}
 \item The process of thermal excitation has a strong non-Markovian component at early times ($\grt \sim \omega^{-1}$) for larger frequencies $\omega \gtrapprox  2 \pi T$.
 \item Relaxation to equilibrium is Markovian, with the thermal decay coefficient $\Gamma =  \Gamma_0   \coth\left(\frac{\grw}{2T}\right)$, except for very low temperatures $T< \frac{\Gamma_0}{2\pi}$, where  relaxation is non-Markovian and the rate is of the order of $T^{-1}$.
 \item  For degenerate qubits $(\omega = 0$), relaxation is strongly non-Markovian and the asymptotic state preserves a large part of the initial state's phase information.
 \end{enumerate}

\subsubsection{Comoving qubit in de Sitter spacetime} 
 We consider a qubit on a comoving geodesic in de Sitter spacetime interacting with a massive scalar field of mass $m$ in its invariant state. At the limit where $m\rightarrow 0$ and the field is conformally coupled to the metric, the positive frequency Wightmann function
   is \cite{Birrell}
\be
\Delta^+(x,x')=-\frac{H^2}{16\pi^2}\frac{1}{\sinh^2\left[H(\grt-\grt'-i\gre)/2\right]-e^{-H(\grt+\grt')}|\mathbf{x} - \mathbf{x}|^2},
\ee
where $\grt$ is the proper time coordinate of comoving observers, $\mathbf{x}$ are spatial coordinates on the surfaces of homogeneity and $H$ is the Hubble constant.
When $\mathbf{x}=\mathbf{x'}=0$
\be\label{dSW}
\Delta^+(\grt; \grt+s)=-\frac{H^2}{16\pi^2}\frac{1}{\sinh^2\left[H(s-i\gre)/2\right]};
\ee
the Wightman function is identical with the correlation function
 (\ref{Wacc}) for a uniformly accelerated qubit or equivalently with a static qubit in a thermal bath at the
 Gibbons-Hawking temperature \cite{Gibbons}
\be
T=\frac{H}{2\pi }.
\ee
The key point of our analysis also passes to the de Sitter case: the thermal behavior is best seen at the long time limit, rather than at the early-time excitation rate, because of the non-Markovian effects.

\section{Conclusions}\label{concl}
The main results of this article are (i) the development of a general method for describing the non-Markovian  dynamics of a qubit interacting with a field bath and (ii) the detailed characterization of the non-Markovian regimes for a uniformly accelerated qubit.

Our approach is particularly suited for studying the open system dynamics for bath environments with static correlation functions.  In this case, we obtain a closed expression for the elements of the reduced density matrix, and the calculation reduces to the implementation of an inverse Laplace transform. In particular, the results presented here can be straightforwardly applied for the study of qubits in rotating motion. We expect that this method can be straightforwardly applied for understanding the non-Markovian dynamics of qubits in  general field baths, including  entanglement dynamics in multi-qubit systems.

Our analysis of the uniformly accelerated qubit  suggests  that the relation between acceleration and temperature is best expressed in terms of the asymptotic state of the detector. Non-Markovian effects are significant at early times, and they may render the early-time transition rate non-thermal. In contrast, the asymptotic state is thermal even when the non-Markovian dynamics is taken into account. Thus, the quantum field vacuum behaves like a thermal reservoir to all accelerated detectors: it brings them to a thermal state, irrespective of their intermediate dynamics.

\begin{acknowledgements}
Research was supported by
Grant No. E611
from the
Research Committee of
the
University
of
Patras
via the "K. Karatheodoris" program.

\end{acknowledgements}

\newpage 
\appendix

\section{Evaluation of the Laplace Transform Integral of Wightman functions}\label{LPT}

 The Laplace transform of the positive frequency Wightman correlation function is
\bea\label{lpint}
\mathcal{L}\left\{\Delta^+(\grt ;0)\right\}(z)&=&-\frac{a^2}{16\pi^2}\int_0^{\infty}e^{-z\grt}\frac{1}{\sinh^2[a(\grt-i\gre)/2]}  d\grt \nn \\
&=&\frac{a}{8\pi^2}\int_0^{\infty}e^{-z\grt}d\left\{\coth[a(\grt-i\gre)/2]\right\}\nn\\
&=&\frac{a}{8\pi^2}\left\{\coth[(ia\gre)/2]+z\int_0^{\infty}e^{-z\grt}\coth[a(\grt-i\gre)/2]d\grt\right\}.
\eea

We need to evaluate the integral
\be\label{LaplInt}
I=\int_0^{\infty}e^{-z\grt}\coth[a(\grt-i\gre)/2]d\grt.
\ee
We calculate the integral (\ref{LaplInt}) as
\bea
I&=&\int_0^{\infty}e^{-z\grt}\left[1-e^{-a(\grt-i\gre)}\right]^{-1}d\grt+\int_0^{\infty}e^{-a(\grt-i\gre)}e^{-z\grt}\left[1-e^{-a(\grt-i\gre)}\right]^{-1}d\grt\nn \\
&=&\frac{1}{a}\int_0^1t^{\frac{z}{a}-1}\left[1-e^{ia\gre}t\right]^{-1}dt+\frac{e^{ia\gre}}{a}\int_0^1t^{\frac{z}{a}}\left[1-e^{ia\gre}t\right]^{-1}dt\nn \\
&=&\frac{1}{z}\ {}_2F_1\left(1,\frac{z}{a},\frac{z}{a}+1;e^{ia\gre}\right)+\frac{e^{ia\gre}}{z+a}\ {}_2F_1\left(1,\frac{z}{a}+1,\frac{z}{a}+2;e^{ia\gre}\right),
\eea
where ${}_2F_1(a,b,c;w)$ is the Gauss hypergeometric function \cite{Abramowitz,Ryzhik}. We used the integral representation of the hypergeometric function
\be
{}_2F_1(a,b,c;w)=\frac{\Gamma(c)}{\Gamma(b)\Gamma(c-b)}\int_0^1 t^{b-1}(1-t)^{c-b-1}(1-tw)^{-a}dt.
\ee
Thus, the Laplace-transformed correlation function (\ref{lpint}) is
\bea\label{LTpos}
\mathcal{L}\left\{\Delta^+(\grt ;0)\right\}(z)=\frac{\gra}{8\pi^2}\Big\{&&\coth[(i\gra\gre)/2]+{}_2F_1\left(1,\frac{z}{\gra},\frac{z}{\gra}+1;e^{i\gra\gre}\right)\nn\\&&+\frac{ze^{ia\gre}}{z+a}\ {}_2F_1\left(1,\frac{z}{\gra}+1,\frac{z}{\gra}+2;e^{i\gra\gre}\right)\Big\}.
\eea
Similarly, we calculate the Laplace transform of the negative frequency Wightman function. We obtain
\bea
\mathcal{L}\left\{\Delta^-(\grt ;0)\right\}(z)=\frac{\gra}{8\pi^2}\Big\{&-&\coth[(i\gra\gre)/2]+{}_2F_1\left(1,\frac{z}{\gra},\frac{z}{\gra}+1;e^{-i\gra\gre}\right)\nn\\&+&\frac{ze^{-ia\gre}}{z+a}\ {}_2F_1\left(1,\frac{z}{\gra}+1,\frac{z}{\gra}+2;e^{-i\gra\gre}\right)\Big\}. \label{LTneg}
\eea

The hypergeomertic series ${}_2F_1(a,b,c;w)$ is analytic everywhere in the complex plane except for the branch points at $w=0,1,\infty$. When $w\to1$, the zero-balanced hypergeometric series, i.e., the series that $c-b-a=0$, behave as
\be
\frac{\Gamma(a)\Gamma(b)}{\Gamma(a+b)}\ {}_2F_1(a,b,a+b;w)=-2\grg-\psi(a)-\psi(b)-\log(1-w)+o(1),
\ee
where $\grg$ is the Euler-Mascheroni constant and $\psi(z)=\frac{d}{dz}\log\Gamma(z)$ is the digamma (psi) function \cite{Bateman,Abramowitz}. Expanding around the branch point $w=1$, we write the Laplaced-transformed correlation functions  (\ref{LTpos}) and (\ref{LTneg}) as
\begin{eqnarray}
\mathcal{L}\left\{\Delta^+(\grt ;0)\right\}(z)=\frac{a}{8\pi^2}\coth\left(\frac{ia\gre}{2}\right)-\frac{a}{8\pi^2}-\frac{z}{4\pi^2}\left[\log(e^{\grg}\gre a)+\grc\left(\frac{z}{a}\right)\right]+\frac{iz}{8\pi},\\
\mathcal{L}\left\{\Delta^-(\grt ;0)\right\}(z)=-\frac{a}{8\pi^2}\coth\left(\frac{ia\gre}{2}\right)-\frac{a}{8\pi^2}-\frac{z}{4\pi^2}\left[\log(e^{\grg}\gre a)+\grc\left(\frac{z}{a}\right)\right]-\frac{iz}{8\pi},
\end{eqnarray}
where we used $\psi(1)=-\gamma$ and $\psi(1+z)=1/z+\psi(z)$. The logarithm is taking values in the principal branch.

\section{Calculation of Markovian poles}\label{Markopolo}
The Laplace-transformed diagonal elements (\ref{r11l}--\ref{r00l}) apparently have a pole at $z=0$. We also find another pole at
\bea
z&=&\frac{g^2a}{2\pi^2}+\frac{g^2}{2\pi^2}i\grw\left[\grc\left(\frac{i\grw}{a}\right)-\grc\left(-\frac{i\grw}{a}\right)\right]+O(g^4)\nn\\
&=&-\frac{g^2\grw}{2\pi}\coth\left(\frac{\pi\grw}{a}\right)+O(g^4)\nn\\
&=&-\Gamma+O(g^4),
\eea
where we used the functional relation $\psi(z)-\psi(-z)=-1/z-\pi \cot(\pi z)$.

The Laplace-transformed off-diagonal element $\tilde{\grr}_{10}(z)$ has two poles, one at
\bea
z&=&\frac{g^2a}{4\pi^2}-\frac{g^2}{2\pi^2}i\grw\left[\log(e^{\grg}\gre a) +\grc\left(-\frac{i\grw}{a}\right)\right]+O(g^4)\nn\\
&=&-\frac{g^2}{2\pi^2}i\grw\left[\log(e^{\grg}\gre\grw)+\log(a/\grw )+Re\left[\grc\left(\frac{i\grw}{a}\right)\right]\right]-\frac{\Gamma}{2}+O(g^4)\nn\\
&=&-i(C+\Delta\grw)-\frac{\Gamma}{2}+O(g^4)
\eea
and  one at
\bea
z&=&2i\grw+\frac{g^2a}{4\pi^2}-\frac{g^2}{2\pi^2}i\grw\left[\log(e^{\grg}\gre a) +\grc\left(\frac{i\grw}{a}\right)\right]+O(g^4)\nn\\
&=&2i\grw-\frac{g^2}{2\pi^2}i\grw\left[\log(e^{\grg}\gre\grw)+\log(a/\grw )+Re\left[\grc\left(\frac{i\grw}{a}\right)\right]\right]-\frac{\Gamma}{2}+O(g^4)\nn\\
&=&2i\grw+i(C+\Delta\grw)-\frac{\Gamma}{2}+O(g^4),
\eea
where we used $Re\,\psi(ix)=Re\,\psi(-ix)$ and $Im\,\psi(ix)=1/(2x)+\frac{\pi}{2}\coth(\pi x)$ with $x\in\mathbb{R}$.

\section{Master equation of a qubit in a thermal bath without the use of RWA}\label{appme}

We derive and solve the second-order master equation for a qubit interacting with a scalar field $\hat{\grf}$ which is in a thermal equilibrium state
\be
\rho_{\grf}=\frac{e^{-\grb H_{\grf}}}{\tr \left(e^{-\grb H_{\grf}}\right)}
\ee
at temperature $T=1/\grb$.
We employ the Born-Markov approximation but not the Rotating Wave Approximation.

The evolution of the qubit's reduced density matrix is given by Eq. (\ref{dmevol}), which is written as
\bea\label{mastereq}
\dot{\hat{\rho}}(\grt)=g^2\int_{0}^{\grt}ds&&\Big\{\Big(\hat{\grs}_-\hat{\grr}(s)\hat{\grs}_-e^{-2i\grw \grt}+\hat{\grs}_-\hat{\grr}(s)\hat{\grs}_+
-\hat{\grs}_+\hat{\grs}_-\hat{\grr}(s)\Big)e^{i\grw(\grt-s)}\nn\\&&+
\Big(\hat{\grs}_+\hat{\grr}(s)\hat{\grs}_+e^{-2i\grw \grt}+\hat{\grs}_+\hat{\grr}(s)\hat{\grs}_-
-\hat{\grs}_-\hat{\grs}_+\hat{\grr}(s)\Big)e^{-i\grw(\grt-s)}\Big\}\text{tr}\Big(\grr_{\grf}(0)\grf(\grt)\grf(s)\Big)\nn\\&&+ \text{h.c.}
\eea
The bath correlation functions are
\be
\text{tr}\Big(\grr_{\grf}(0)\grf(\grt)\grf(s)\Big)=\int\frac{d^3\mathbf{k}}{(2\pi)^32\grw_{\mathbf{k}}}\left[e^{-i\grw_{\mathbf{k}}(\grt-s)}\Big(n(\grw_{\mathbf{k}})+1\Big)+e^{i\grw_{\mathbf{k}}(\grt-s)}n(\grw_{\mathbf{k}})\right],
\ee
where
\be
n(\grw_{\mathbf{k}})=\frac{1}{e^{\grb\grw_{\mathbf{k}}}-1}
\ee
is the Planck distribution. We apply the Markov approximation and evaluate the time integrals in (\ref{mastereq}) using the formula
\be
\int_0^{\infty}dx e^{\pm i\gre x}=\pi \grd(\gre)\pm PV \frac{1}{\gre},
\ee
where $PV$ indicates the Cauchy principal value. We then transform back to the Schr\"odinger picture to obtain the second-order master equation
\bea\label{master}
\dot{\hat{\rho}}(\grt)=&&-\frac{i}{2}(\grw+\Delta\grw+\Delta'\grw)[\hat{\grs_3},\hat{\grr}(\grt)]-(\Delta\grw+\Delta'\grw)\left(\hat{\grs}_+\hat{\grr}(\grt)\hat{\grs}_+-\hat{\grs}_-\hat{\grr}(\grt)\hat{\grs}_-\right)\nn\\&&+\frac{\Gamma_0}{2}(n(\grw)+1)\left(2\hat{\grs}_-\hat{\grr}(\grt)\hat{\grs}_+-\hat{\grs}_+
\hat{\grs}_-\hat{\grr}(\grt)-\hat{\grr}(\grt)\hat{\grs}_+\hat{\grs}_-+\hat{\grs}_-\hat{\grr}(\grt)\hat{\grs}_-+\hat{\grs}_+\hat{\grr}(\grt)\hat{\grs}_+
\right)\nn\\
&&+\frac{\Gamma_0}{2}n(\grw)\left(2\hat{\grs}_+\hat{\grr}(\grt)\hat{\grs}_--\hat{\grs}_-
\hat{\grs}_+\hat{\grr}(\grt)-\hat{\grr}(\grt)\hat{\grs}_-\hat{\grs}_++\hat{\grs}_-\hat{\grr}(\grt)\hat{\grs}_-+\hat{\grs}_+\hat{\grr}(\grt)\hat{\grs}_+\right),
\eea
where
\be
\Gamma_0=2\pi g^2\int\frac{d^3\mathbf{k}}{(2\pi)^32\grw_{\mathbf{k}}}\left[\grd(\grw_{\mathbf{k}}-\grw)-
\grd(\grw_{\mathbf{k}}+\grw)\right]=\frac{g^2\grw}{2\pi}
\ee
is the zero temperature spontaneous emission rate,
\be
\Delta\grw=g^2 PV \int\frac{d^3\mathbf{k}}{(2\pi)^32\grw_{\mathbf{k}}}\left(\frac{1}{\grw_{\mathbf{k}}+\grw}-\frac{1}{\grw_{\mathbf{k}}-\grw}\right)
\ee
is a frequency renormalization term  and
\be
\Delta'\grw=g^2 PV \int\frac{d^3\mathbf{k}}{(2\pi)^3\grw_{\mathbf{k}}}\left(\frac{1}{\grw_{\mathbf{k}}+\grw}-\frac{1}{\grw_{\mathbf{k}}-\grw}\right)n(\grw_{\mathbf{k}})
\ee
is the temperature-dependent Lamb shift. Writing the density operator in its matrix form (\ref{dmatrix}) and taking the Laplace transform, we obtain
\bea
z\tilde{\grr}_{11}(z)-\grr_{11}(0)&=&-\Gamma_0(n(\grw)+1)\tilde{\grr}_{11}(z)+\Gamma_0n(\grw)\tilde{\grr}_{00}(z)\label{app11}\\
z\tilde{\grr}_{00}(z)-\grr_{00}(0)&=&-\Gamma_0n(\grw)\tilde{\grr}_{00}(z)+\Gamma_0(n(\grw)+1)\tilde{\grr}_{11}(z)\\
z\tilde{\grr}_{10}(z)-\grr_{10}(0)&=&-\left[i\bar{\grw}+\frac{\Gamma_0}{2}(2n(\grw)+1)\right]\tilde{\grr}_{10}(z)\nn\\&&-\left[i\left(\Delta\grw+\Delta'\grw\right)-\frac{\Gamma_0}{2}(2n(\grw)+1)\right]\tilde{\grr}_{01}(z)\label{app10}
\eea
where $\bar{\grw}=\grw+\Delta\grw+\Delta'\grw$ and $\tilde{\grr}_{mn}(z)=\mathcal{L}\{\grr_{mn}(\grt)\}(z)$ is the Laplace-transformed density matrix elements. We solve Eqs. (\ref{app11})--(\ref{app10}) for $\tilde{\grr}_{mn}(z)$
\be
\tilde{\grr}_{11}(z)=\frac{z+\Gamma_0n(\grw)}{z(z+\Gamma)}\grr_{11}(0)+\frac{\Gamma_0n(\grw)}{z(z+\Gamma)}\grr_{00}(0),
\ee
\be
\tilde{\grr}_{10}(z)=\frac{\left(z-i\bar{\grw}+\frac{\Gamma}{2}\right)\grr_{10}(0)-\left[i\left(\Delta\grw+\Delta'\grw\right)-\frac{\Gamma}{2}\right]\grr_{01}(0)}{z^2+\Gamma z+\bar{\grw}^2-\left(\Delta\grw+\Delta'\grw\right)^2},
\ee
where
\be
\Gamma=\Gamma_0(2n(\grw)+1)=\Gamma_0\coth(\grb\grw/2)
\ee
is the thermal decay constant.

The diagonal elements have two poles, one at $z = 0$ and one at $z = - \Gamma$. We calculate the Bromwich integral to obtain
\bea
\grr_{11}(\grt)&=&\frac{1}{2}\left(1-\frac{\Gamma_0}{\Gamma}\right)+\frac{\Gamma_0}{2\Gamma}e^{-\Gamma \grt}-\frac{1}{2} e^{-\Gamma \grt}  (\grr_{00}(0)-\grr_{11}(0)), \\
\grr_{00}(\grt) &=& 1 - \grr_{11}(\grt).
\eea

The off-diagonal elements have two poles at $z=-\frac{\Gamma}{2} \pm i\bar{\grw}$ (in order to find the poles we assumed that $\grg<<\bar{\grw}$). We calculate the Bromwich integral to obtain
\be
\rho_{10}(\grt)=e^{-i\bar{\grw}\grt-\frac{\Gamma}{2}\grt}\grr_{10}(0)+\left[\frac{\Gamma}{2}-i(\bar{\grw}-\grw)\right]\frac{\sin(\bar{\grw}\grt)}{\bar{\grw}}e^{-\frac{\Gamma}{2}\grt}\grr_{01}(0).
\ee
The first term of $\rho_{10}(\grt)$ is the usual one obtained from the second-order master equation, if both the Born-Markov and the Rotation Wave approximation are used \cite{BrePe07,Carmichael}. The second term appears when one does not neglect the counter-rotating terms \cite{Agarwal, RWA}.

\end{document}